# Heat capacity and magnetoresistance in $Dy(Co,Si)_2$ compounds

Niraj K. Singh[a], K. G. Suresh[a], A. K. Nigam[b] and S. K. Malik[b]
[a]*Department of Physics, Indian Institute of Technology Bombay, Mumbai- 400 076, India*
[b]*Tata Institute of Fundamental Research, Homi Bhabha Road, Mumbai – 400 005, India*

**Abstract:**

Magnetocaloric effect and magnetoresistance have been studied in $Dy(Co_{1-x}Si_x)_2$ [x=0, 0.075 and 0.15] compounds. Magnetocaloric effect has been calculated in terms of adiabatic temperatue change ($\Delta T_{ad}$) as well as isothermal magnetic entropy change ($\Delta S_M$) using the heat capacity data. The maximum values of $\Delta S_M$ and $\Delta T_{ad}$ for $DyCo_2$ are found to be 11.4 $JKg^{-1}K^{-1}$ and 5.4 K, respectively. Both $\Delta S_M$ and $\Delta T_{ad}$ decrease with Si concentration, reaching a value of 5.4 $JKg^{-1}K^{-1}$ and 3 K, respectively for x=0.15. The maximum magnetoresistance is found to about 32% in $DyCo_2$, which decreases with increase in Si. These variations are explained on the basis of itinerant electron metamagnetism occurring in these compounds.

## I. INTRODUCTION

The variety of magnetic phenomena exhibited by many rare earth (R) - transition metal (TM) intermetallic compounds render them suitable for applications specially those based on magnetocaloric effect (MCE) and magnetoresistance (MR)[1,2]. Among the R-TM intermetallics, $RCo_2$ compounds have attracted a lot of attention owing to the first order magnetic transition (FOT) in compounds with R= Dy, Ho and Er. The occurrence of FOT in these compounds leads to large MCE and MR, making them promising candidates for applications based on magnetic refrigeration and MR. The application of any magnetic material as magnetic refrigerant requires considerable MCE over large span of temperature and hence various substitutions at rare earth and cobalt site have been studied in these compounds [2,3]. Recently, some of us have reported[4] the MCE in $Dy(Co_{1-x}Si_x)_2$ in terms of isothermal magnetic entropy change ($\Delta S_M$), on the basis of magnetization data. In this paper, we report the MCE of these compounds in terms of adiabatic temperature change ($\Delta T_{ad}$) using the heat capacity (C) data. We also calculate the $\Delta S_M$ values using the same data and compare it with that obtained from the magnetization isotherms. In addition, we have also studied the variation of MR in these compounds.

## II. EXPERIMENTAL DETAILS

All the compounds [x=0, 0.075 and 0.15] were prepared and characterized by methods reported elsewhere[4]. The heat capacity and electrical resistivity were measured in the temperature range 2- 240 K and in fields up to 50 kOe, using a PPMS (Quantum Design). The heat capacity was measured using the relaxation method and the electrical resistivity was measured employing the linear four-probe technique.

## III. RESULTS AND DISCUSSION

All the compounds possess the cubic $MgCu_2$ structure. The ordering temperatures ($T_C$) of the compounds with x= 0, 0.075 and 0.15 are 139, 164 and 154 K, respectively[4]. Fig.1 shows the temperature variation of heat capacity for $DyCo_2$ in zero field and in a field of 50 kOe. Similar plots were obtained for Si substituted compounds as well. It can be seen from the figure that the zero field heat capacity shows a maximum near $T_C$. The $T_C$ values determined from zero field heat capacity data are in close agreement with that calculated from magnetization measurements. The presence of zero field heat capacity peak is attributed to the absorption of large amount of heat, which is utilized in randomizing the magnetic moments near $T_C$. In the presence of field, the process of randomization of moments takes place at temperatures above $T_C$, thereby broadening the peak.

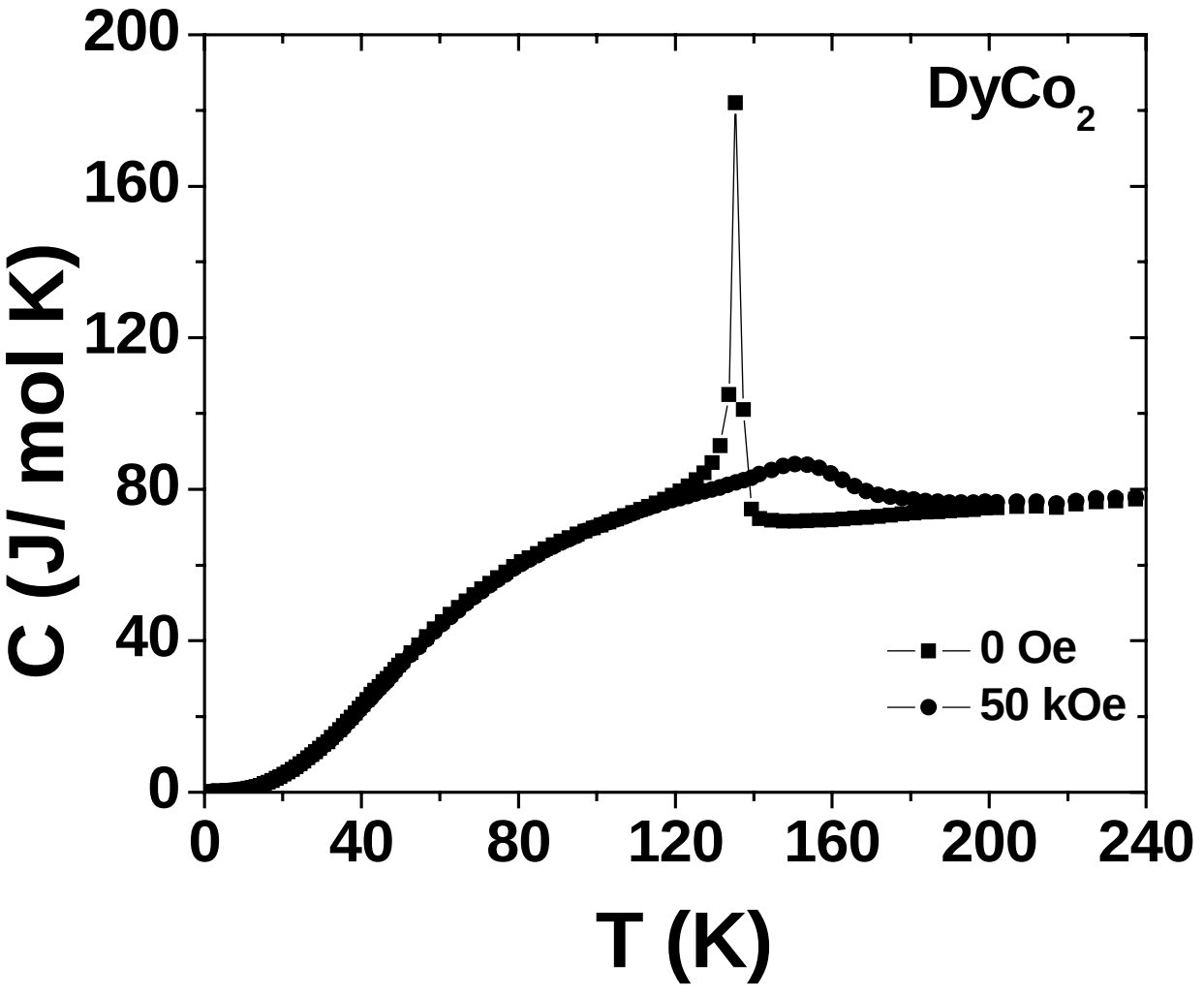

FIG. 1 The temperature dependence of heat capacity of $DyCo_2$ in zero field and at 50 kOe.

The $\Delta S_M$ and $\Delta T_{ad}$ have been calculated using the heat capacity data measured in different fields employing the method reported in literature[5]. The variation of $\Delta S_M$ and $\Delta T_{ad}$ for all the compounds for a field change of 50 kOe is shown in Fig. 2a and 2b, respectively along with the results obtained from magnetization isotherms. It can be seen from Fig. 2a that the $\Delta S_M$ calculated from heat capacity data compares very well with that calculated from the magnetization isotherms[4].

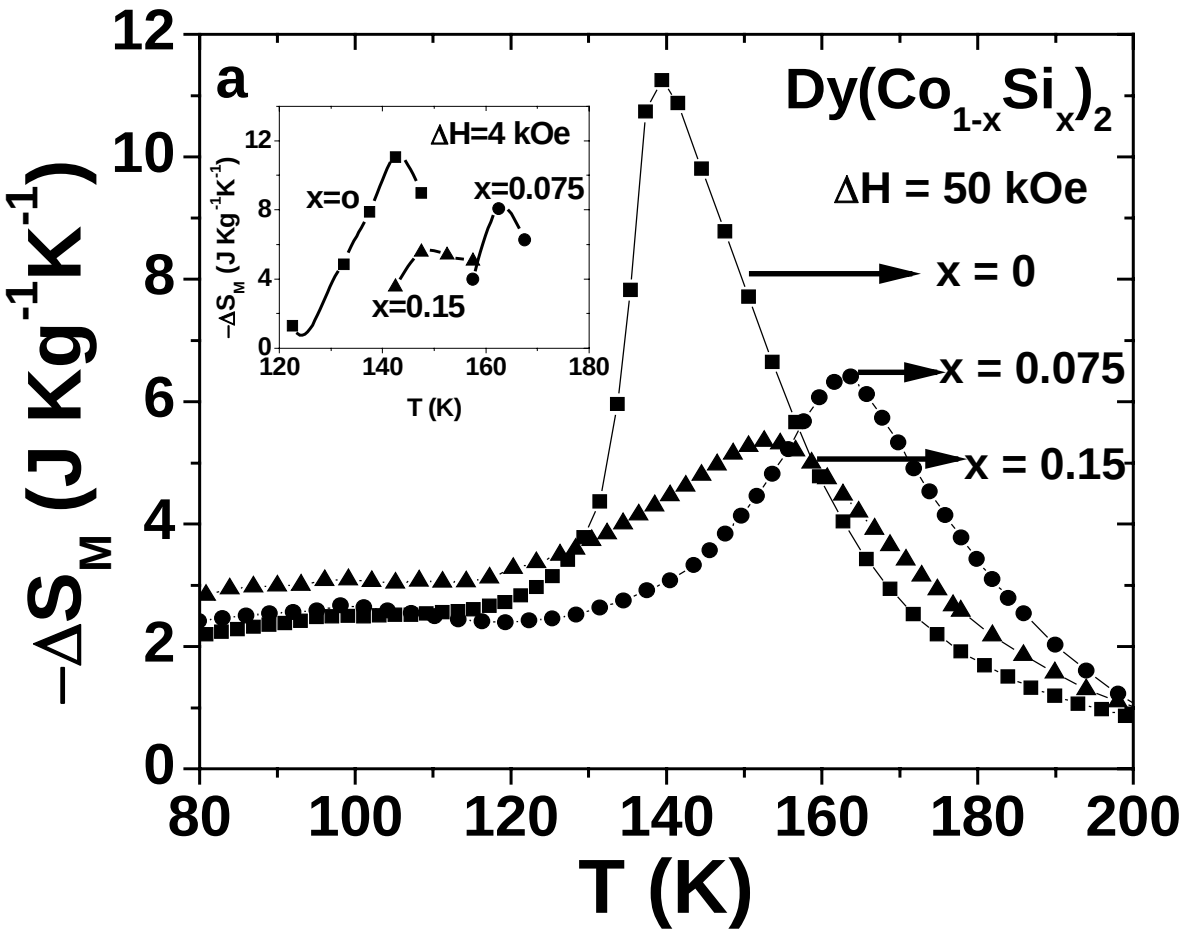

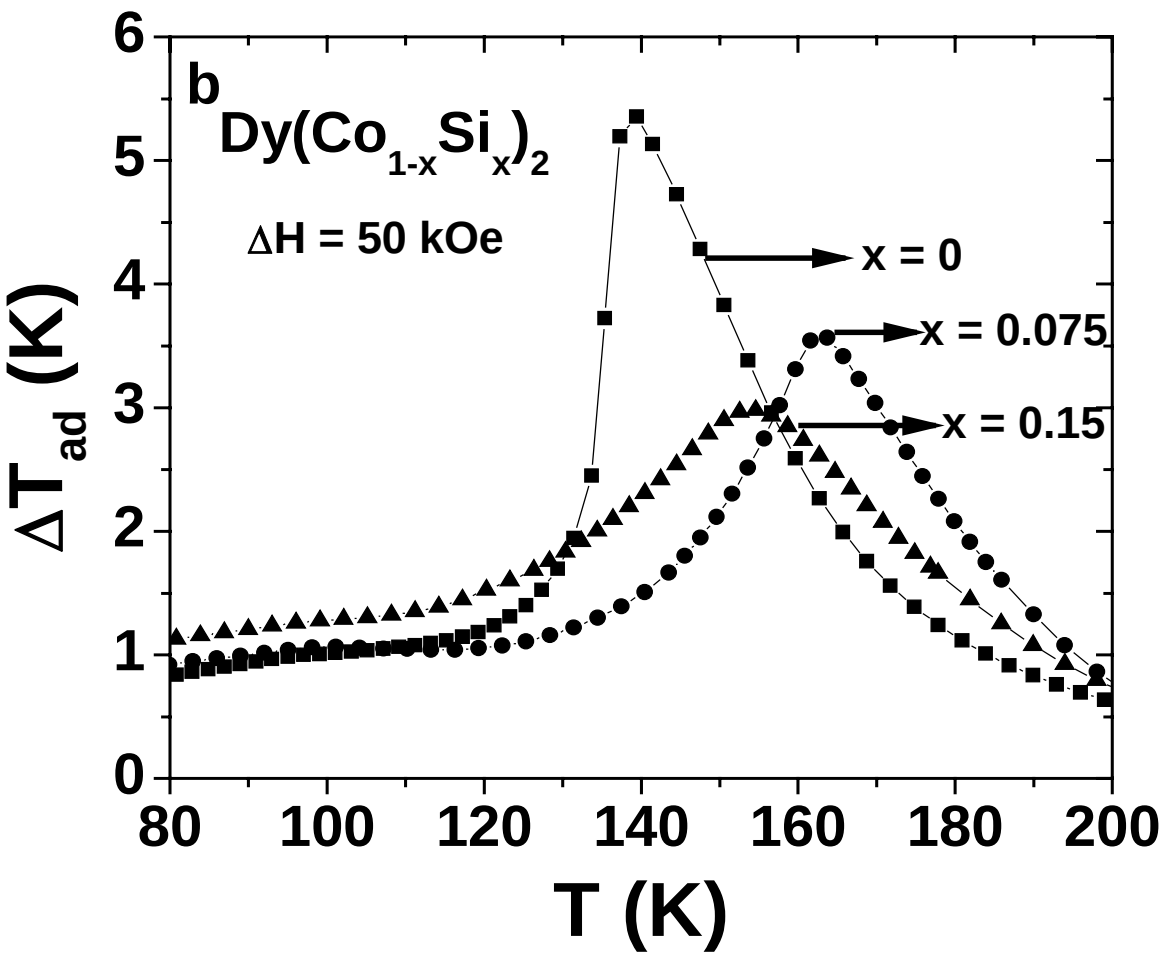

FIG. 2 Temperature dependence of (a) $\Delta S_M$ and (b) $\Delta T_{ad}$, for $\Delta H$= 50 kOe, in $Dy(Co_{1-x}Si_x)_2$ compounds. The inset in 2a shows the temperature dependence of $\Delta S_M$ calculated from the magnetization data for $\Delta H$= 40 kOe.

It can be seen from the figures that both $\Delta S_M$ and $\Delta T_{ad}$ show a peak at $T_C$ and also the peak height decreases with increase in Si concentration. The maximum value of $\Delta S_M$ and $\Delta T_{ad}$ ($\Delta S_M^{max}$ and $\Delta T_{ad}^{max}$ ) in $DyCo_2$ for $\Delta H$= 50 kOe are 11.4 $JKg^{-1}K^{-1}$ and 5.4 K, respectively. However, the $\Delta S_M^{max}$ and $\Delta T_{ad}^{max}$ in $Dy(Co_{0.85}Si_{0.15})_2$ decrease to 5.4 $JKg^{-1}K^{-1}$ and 3 K, respectively. The large value of MCE in $DyCo_2$ is attributed to the occurrence FOT at $T_C$. The creation of Co moments due to the molecular field of Dy, as the sample is cooled through $T_C$, is known as itinerant electron metamagnetism (IEM) and is responsible for FOT[6]. The reduction in MCE with Si is mainly due to the weakening of IEM as a result of spin fluctuations[4].

The MCE values obtained in the present system are comparable to those of Gd, which is considered as a potential refrigerant[5]. Furthermore, it is of interest to note that the MCE is significant over the temperature range 130-170 K, which makes this system a promising candidate as a composite material for magnetic refrigeration applications near 150 K.

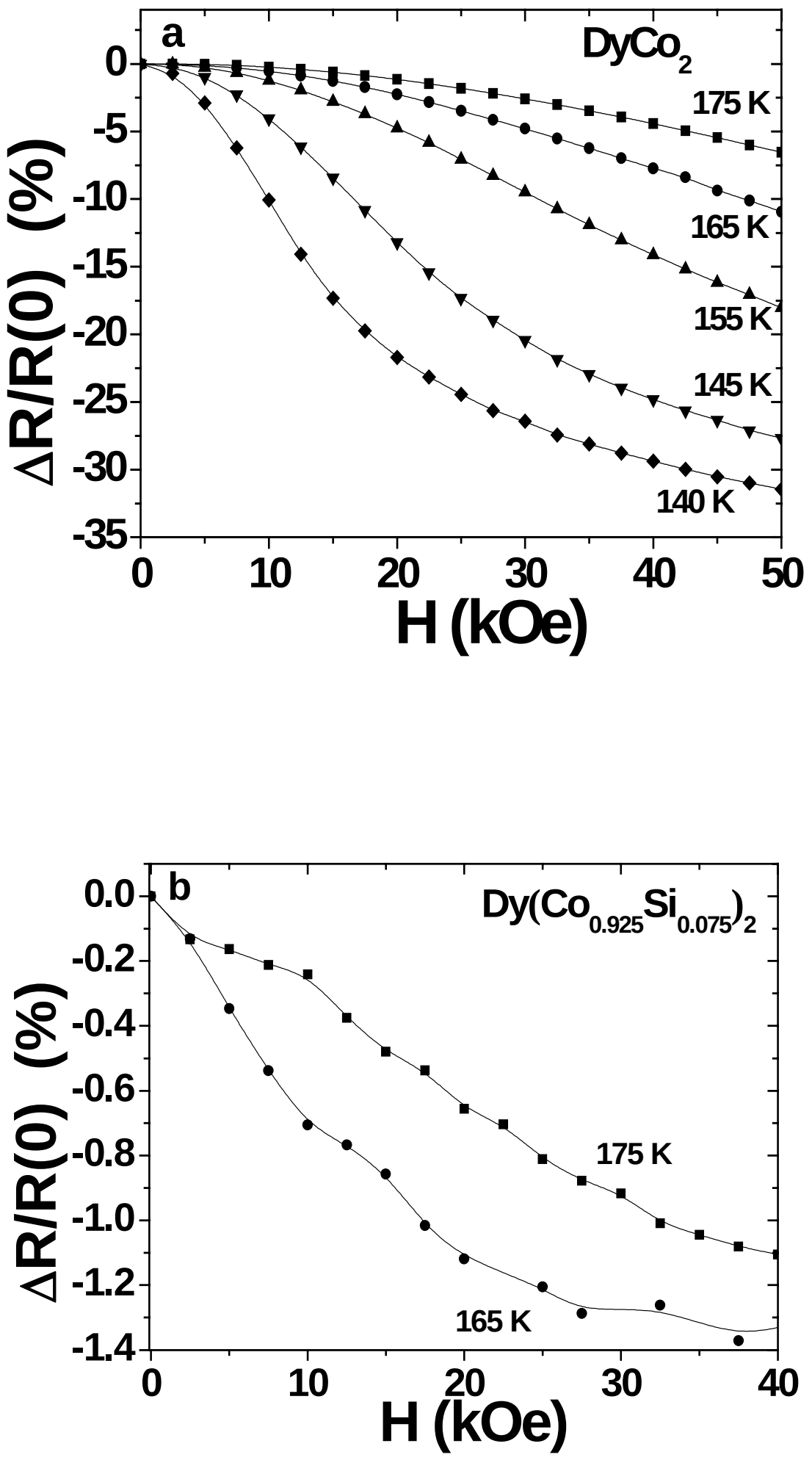

FIG. 3 Variation of magnetoresistance with applied field in (a) $DyCo_2$ and (b) $Dy(Co_{0.925}Si_{0.075})_2$ at various temperatures.

The electrical resistivity has been measured under various fields starting from zero to a maximum of 50 kOe. The temperature variation of the zero-field resistivity data shows that all the samples are metallic. The residual resistivity increases with Si concentration, which is due to the additional scattering centers as a result of Si substitution. The magnetoresistance ratio [*ΔR/R(0)*] where *ΔR* is the change in the resistivity and *R(0)* is the zero-field resistivity has been calculated for all the compounds. Figs. 3a and b show the variation of magnetoresistance ratio as a function of applied field at various temperatures in $DyCo_2$ and $Dy(Co_{0.925}Si_{0.075})_2$, respectively.

It can be seen from the Fig. 3a that the MR of $DyCo_2$, just above $T_C$, is negative and the value is about 32% for a field of 50 kOe, which is in close agreement with that reported earlier[2]. With increase of temperature, the MR is found to decrease, still retaining the negative sign. In Si substituted compounds, the MR is found to be much smaller as compared to that in $DyCo_2$, as shown in Fig. 3b. The variations of MR with temperature and Si concentration can be analyzed as follows.

The change in the electrical resistivity caused by the application of a magnetic field in the case $RCo_2$ compounds can be written as [2,7]

$$\Delta R\,(H,T) = \Delta R_{3d}\,(H,T) + \Delta R_{4f}\,(H,T) + \Delta R_{sf}\,(H,T) + \Delta R_L\,(H,T) \qquad (1)$$

where the first term on the right side represents the contribution due to the 3d moment formation. Since this leads to additional magnetic scattering $\Delta R_{3d}\,(H,T)$ is positive. The origins of this term are the molecular field of the rare earth and the magnetovolume effect. The second and the third terms are due to the 4f –spin disorder scattering and the spin fluctuations, respectively. The spin disorder scattering represents the extra scattering that the conduction electrons suffer as the material enters the paramagnetic phase. The spin fluctuations arise both in the 4f sublattice as well as in the 3d sublattice. In R-TM systems, where both R and TM ions possess permanent moments, the 3d sublattice contribution to the spin fluctuations is more than that of 4f sublattice due to the itinerant nature of the former. The applied magnetic field suppresses both spin disorder and spin fluctuations, rendering these two contributions negative. The last term is due to the Lorentz force produced by the applied field on the conduction electrons and this term is also positive. Gratz et al.[7] have recently reported that in spin fluctuating $RCo_2$ systems, the positive MR due to this term is temperature dependent. This contribution is typically about 5% at temperatures above 100 K for a field of 50 kOe. It can be seen from Fig. 3a that in $DyCo_2$ the predominant contribution towards $\Delta R$ is negative, which implies that the negative contributions are stronger than the positive ones. The formation of Co moments in $DyCo_2$ with the help of Dy molecular field as the temperature decreases towards $T_C$ is the main source of $\Delta R_{3d}\,(H,T)$. The 3d moment formation is possible only at temperatures ≤ $T_C$. It is of interest to note that by virtue of the smaller lattice parameter of $DyCo_2$, the existence of 3d moments above $T_C$ is very less probable[4]. As $T_C$ is approached from the high temperature region, the IEM causes a strong suppression of spin fluctuations at $T_C$ and hence results in a large negative MR. At temperatures well above $T_C$, the applied field becomes less effective in suppressing the spin disorder and spin fluctuations and hence the MR decreases. Since the suppression of electrical resistivity with field is quite high near $T_C$, the Lorentz force contribution becomes quite negligible in $DyCo_2$.

In the Si- substituted compound, the MR just above $T_C$ is quite small. It is reasonable to assume that $\Delta R_{4f}$ *(H,T)* contribution to MR is almost the same as in $DyCo_2$. With Si substitution, the lattice parameter increases towards the critical value needed for the 3d moment formation. As a result of this, 3d moments may exist at temperatures above $T_C$ also. Therefore, the change in the electrical resistivity as the sample enters the ferromagnetic phase may not be as abrupt as in $DyCo_2$. Furthermore, the positive contribution from $\Delta R_L(H,T)$ term may become significant in this case, thereby reducing the negative MR considerably in Si substituted compounds. With increase in temperature, the MR drops as in the case of $DyCo_2$. In the case of $Dy(Co_{0.85}Si_{0.15})_2$, the MR tends to become positive though the value is quite small.

## IV. CONCLUSION

In conclusion, $Dy(Co,Si)_2$ system shows considerable MCE in the temperature range 130-170 K, which makes this system attractive from the point of view of magnetic refrigeration applications. The dependence of magnetoresistance on the Si concentration is found to be similar to that of MCE. The IEM appears to play a significant role in the variation of both MCE and MR in these compounds.

**ACKNOWLEDGEMENTS**

One of the authors (KGS) thanks the D.S.T., Govt. of India for financial support in the form of a sponsored project. The authors acknowledge the help rendered by Dr. R. Nirmala and D. Buddhikot in carrying out the measurements.

---

[1]K. A. Gschneidner Jr. and V. K. Pecharsky, Mater. Science and Engineering **A287**, (2000) 301

[2]N. H. Duc, D. T. Kim Anh and P. E. Brommer, Physica **B 319** (2002) 1

[3]D. Wang, H. Liu, S. Tang, S. Yang, S. Huang and Y. Du, Physics Lett. **A 297** (2002) 247

[4]Niraj K. Singh, K. G. Suresh and A. K. Nigam, Solid State Commun. **127** (2003) 373

[5]V. K. Pecharsky and K. A. Gschneidner, Jr., J. Magn. Magn. Mater. **200**, (1999) 44

[6]S. Khmelevskyi and P. Mohn, J. Phys. Cond. Matter. **12** (2000) 9453

[7]E. Gratz, H. Nowotny, J. Enser, E. Bauer and K. Hense, J. Magn. Magn. Mater. **272-276**, (2004) c441